\documentclass[aps, pra, twocolumn, nofootinbib, notitlepage, superscriptaddress   
]{revtex4-1}

\usepackage{amssymb,latexsym,amsmath}
\usepackage{bm}
\usepackage{amssymb,amsmath}
\usepackage{graphicx,xcolor}

\usepackage{hyperref}   
\hypersetup{%
	pdfpagemode=None, 
	pdfstartpage=1,
	pdfmenubar=true,
	pdftoolbar=true,
	colorlinks = true,
	linkcolor=blue,
	citecolor=blue,
	urlcolor=blue,
	bookmarksopen=false
}

\newcommand{\beq}{\begin{equation}}   
\newcommand{\eeq}{\end{equation}}

\newcommand{\br}{{\bm r}}

\begin{document}

\title{Superfluid Vortex Dynamics on Planar Sectors and Cones}
\author {Pietro Massignan}
\email{pietro.massignan@upc.edu}
\affiliation{Departament de F\'isica, Universitat Polit\`ecnica de Catalunya, E-08034 Barcelona, Spain}
\affiliation{ICFO -- Institut de Ci\`encies Fot\`oniques, The Barcelona Institute of Science and Technology, 08860 Castelldefels (Barcelona), Spain}
\author {Alexander L.\ Fetter}
\email{fetter@stanford.edu}
\affiliation {Departments of Physics and Applied Physics, Stanford University, Stanford, CA 94305-4045, USA}
\date{\today}

\begin{abstract}

We study the dynamics of vortices formed in a superfluid film adsorbed on the curved two-dimensional surface of a cone. To this aim, we observe  that a cone can be unrolled to a sector on a plane with periodic boundary conditions on the straight sides.  The sector can then be mapped conformally to the whole plane, leading to the relevant stream function.  In this way, we show that a superfluid vortex on the cone precesses uniformly at fixed distance from the apex.
The stream function also yields directly the interaction energy of two vortices on the cone.  We then study the vortex dynamics on unbounded and bounded cones.
In suitable limits, we recover the known results for dynamics on cylinders and planar annuli.
\end{abstract}

\maketitle

\section{Introduction}\label{sec:Intro}
Vortex dynamics in superfluid films depends strongly on the shape of the underlying surface. For example,  a region with local positive Gaussian curvature (such as the top of a smooth hill, or the bottom of a valley) exerts a repulsive force on a vortex~\cite{Vite04,Turn10}. Geometries with vanishing Gaussian curvature also have special features:  for example, single vortices on an infinite cylinder have quantized azimuthal velocities \cite{Guen17} because of the single-valued nature of the condensate wave function.  Here, we study the different and interesting case of superfluid vortex dynamics on a conical surface, which is equivalent to the motion on a planar sector~\cite{Schi16}.

We consider a  sector (a wedge) of the plane with opening angle $2\pi/\alpha$, where $\alpha$ is real and larger than one.  We also impose periodic boundary conditions on the two radial sides.
To determine the  hydrodynamic flow arising from a singly quantized  vortex at some complex position $z_0$ in the sector, we use the conformal transformation $Z(z) = z^\alpha$ from the sector to the whole plane.  Reference~\cite{Schi16} considered the special case $\alpha = 3$, but this transformation holds for general  $\alpha \ge 1$~\cite{Rile06}.  Note that the limiting case $\alpha \to 1 $ represents the whole plane.

Section \ref{sec:cone} provides basic background for this problem, including the complex potential and the self-induced motion associated with the curved surface of the cone.  The next Secs.~\ref{sec:energy} and~\ref{dyn}  consider the energy and dynamics of a system of vortices, where the interaction energy can be expressed simply in terms of the  stream function. Section \ref{sec:finiteCones} generalizes to the case of bounded cones, both internally and externally.  The last Sec.~\ref{sec:conclusions} concludes with discussion and a brief conjecture about the situation for $0 < \alpha < 1$.

\section{Vortices on sectors and cones}\label{sec:cone}

We start from the full plane with complex coordinate $Z = X + i Y $.
A fundamental tool for the description of hydrodynamics of two-dimensional  incompressible and irrotational fluids is the complex potential $F(Z) = \chi + i \Phi$, where $\chi$ is the stream function and $\Phi$ is the velocity potential.
Either function provides the hydrodynamic flow pattern with the general relation $v_y + i v_x =(\hbar/M) F'(Z)$, with $M$ the mass of the superfluid particles.
For a singly quantized positive vortex at $Z_0 = X_0 + i Y_0$, the complex potential is 
\beq\label{Fplane}
F_{\rm plane}(Z)=\ln(Z-Z_0),
\eeq
 and we  use a conformal map to obtain the corresponding complex potential on the sector and the cone. 

\subsection{Geometry of cones and sectors} 
It is clear from elementary considerations that a finite cone (like a ``dunce cap" or a ``witch hat") may be unrolled onto a wedge-shaped sector. Similarly, an infinite cone may be unfolded onto an infinite sector of the plane, or a truncated cone unfolds to a portion of an annulus. This procedure does not introduce any distortion, and therefore preserves both lengths and areas. 
It is also conformal, since it preserves angles locally.

Let us here briefly review the geometric descriptions of a cone, and of the corresponding sector.  The usual spherical polar coordinates ($r,\theta,\phi$) can make this connection precise, since an unbounded cone is the axisymmetric surface associated with  fixed polar angle $\theta$, leaving $r$ and $\phi$ as the two parameters specifying locations on the surface.   In general,  the three-dimensional coordinate vector becomes 
\begin{equation}
\bm r = r\sin\theta\cos\phi\,\hat{\bm x} + r\sin\theta\sin\phi\,\hat{\bm y} + r\cos\theta\, \hat{\bm z}
\end{equation}
expressed in spherical polar coordinates.  The  partial derivatives $\partial\bm r/\partial r$, $\partial \bm r/\partial \theta$, and $\partial \bm r/\partial \phi$ yield (after normalization) the three orthogonal unit vectors $\hat{\bm r}$, $\hat{\bm \theta}$, and $\hat{\bm \phi}$  (see, for example Sec.~10.9.2 in Ref.~\cite{Rile06}).  On the surface of the cone, the  outward unit normal is $\hat{\bm n} = \pm \hat{\bm \theta}$ with $\pm 1 = {\rm sgn}(\cos\theta) =\cos\theta/|\cos\theta| $.

\begin{figure}
\begin{center}
\includegraphics[width=.2\linewidth]{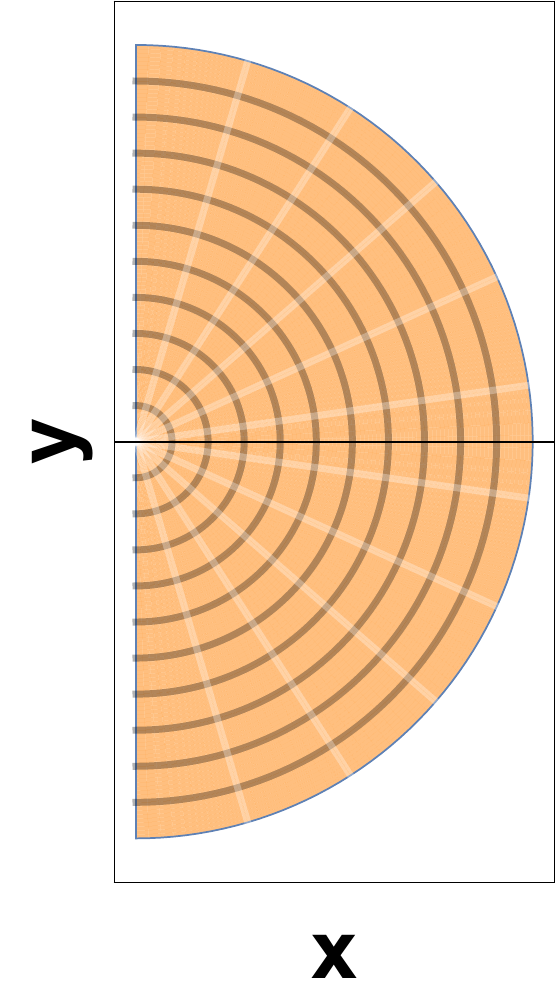}\hspace{1cm}
\includegraphics[width=.48\linewidth]{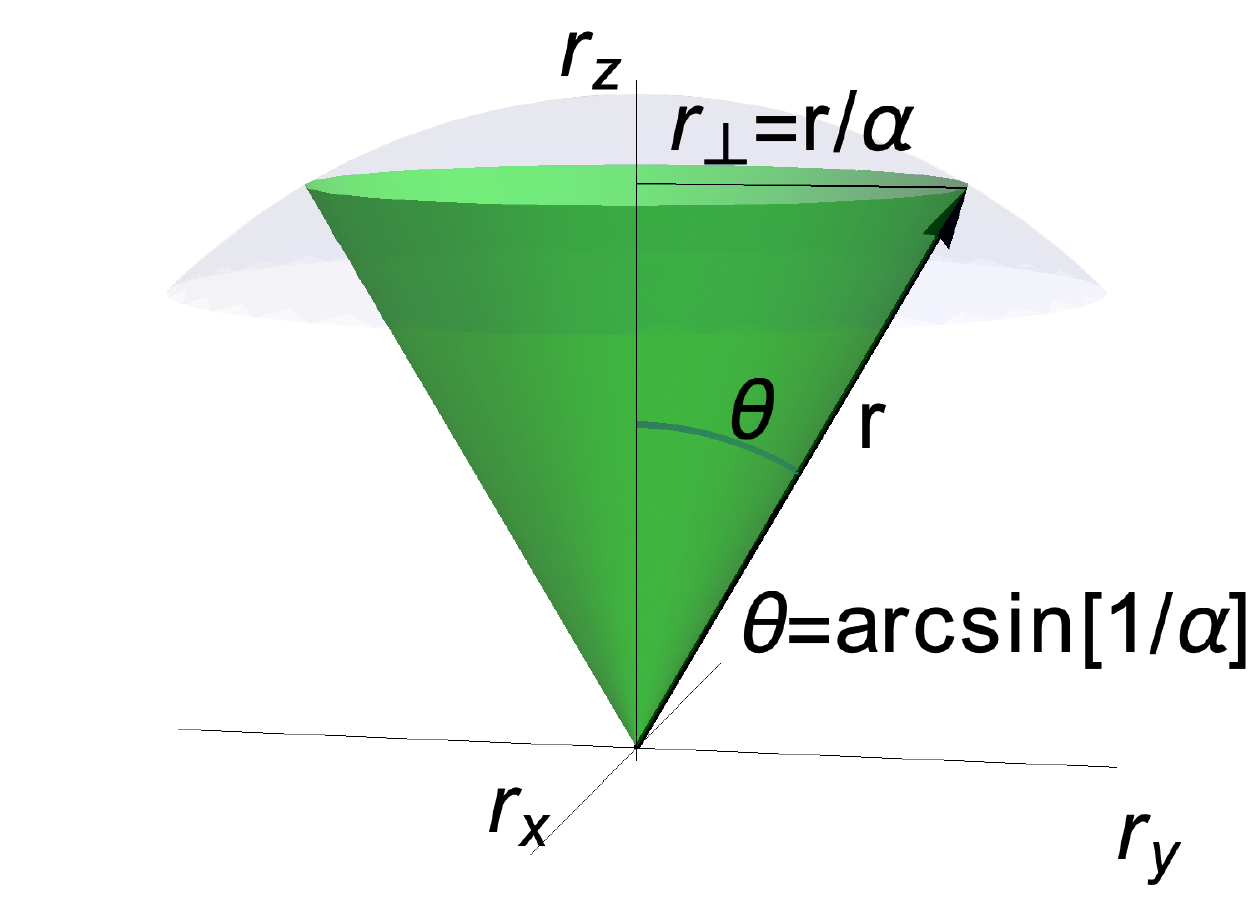}
\caption{\label{fig:geometry_of_cone}
A planar sector with opening angle $2\pi/\alpha$ (left panel) may be wrapped up to form a cone with aperture angle $\theta=\arcsin(1/\alpha)$ (right panel). Here, we choose $\alpha=2$.}
  \end{center}
 \end{figure}

The apex angle of the cone is $\theta$ for $\theta < \pi/2$ and $\pi - \theta$ for $\theta > \pi/2$, and the perpendicular distance from the symmetry axis to the conical surface is $r_\perp = r\sin\theta$.  Let $\sin\theta =1/\alpha$ or, equivalently $\theta = \arcsin(1/\alpha)$, both with $\alpha \ge 1$.  Hence the perpendicular distance becomes $r_\perp = r/\alpha$, and $\cos\theta = \sqrt{\alpha^2-1}/\alpha$. An illustrative sketch of the geometry under consideration is given in Fig.~\ref{fig:geometry_of_cone}.

\begin{figure*}
\begin{center}
\includegraphics[width=\linewidth]{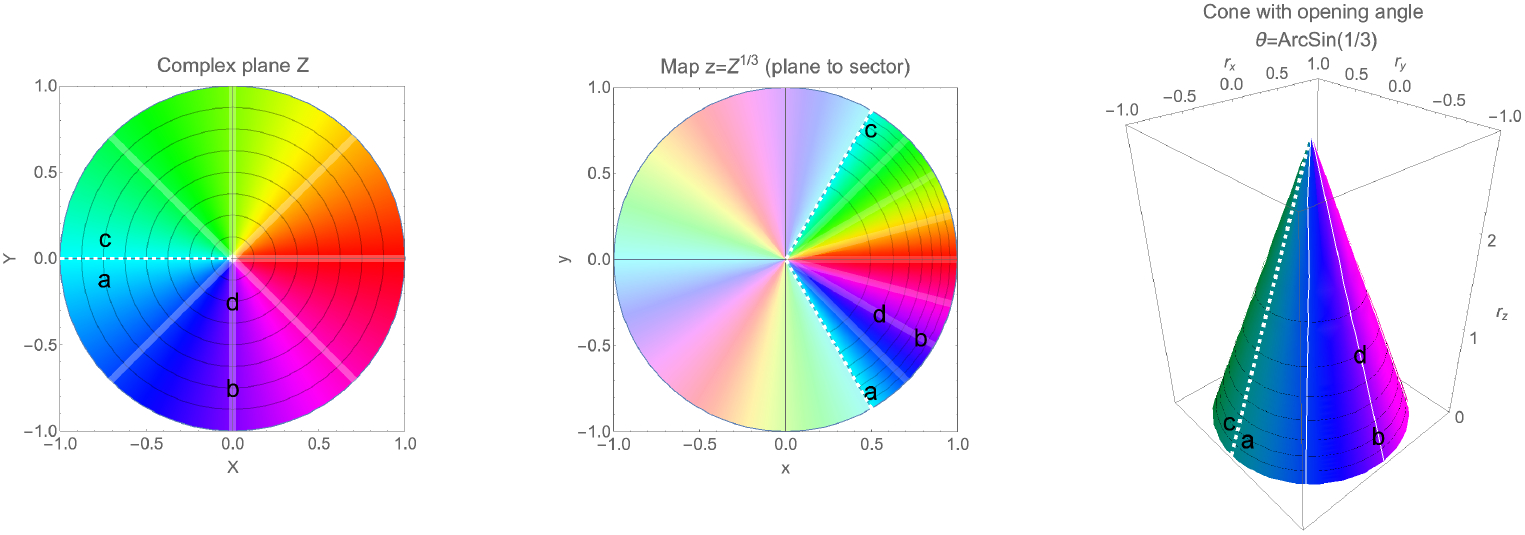}
\caption{\label{fig:map_to_cone}
{\bf Conformal map to a sector and cone.}  
The map $z=Z^{1/\alpha}$ with $\alpha\geq1$ sends the complex plane $Z$ (left) onto a sector with opening angle $2\pi/\alpha$ (center). This sector may be folded onto a three-dimensional cone with opening angle $\theta=\arcsin(1/\alpha)$ (right). The images show the specific case $\alpha=3$. The Hue coloring of the maps corresponds to the azimuthal angle of the coordinate on the source plane, $\phi={\rm Arg}(Z)$.
}
  \end{center}
 \end{figure*}

On the surface of the cone, the element of distance is $d\bm s = \hat{\bm r} \,dr + \hat{\bm \phi}\,r_\perp d\phi$. Introducing the parameter $ \bar\phi\equiv \phi/\alpha $, which has the range $0\le \bar\phi \le 2\pi/\alpha$, the element of distance becomes $ d\bm s = \hat{\bm r} \,dr + \hat{\bm \phi}\,r\, d\bar \phi   $, which can now be treated as  the element of distance on an unbounded  planar  sector (or wedge) of angular opening $2\pi/\alpha$.
 In this way, we have a direct mapping from the surface of the cone with spherical polar coordinates $r,\theta,\phi$ and fixed $\theta$ to the planar sector with plane polar coordinates $r, \bar\phi$ and wedge angle $2\pi/\alpha$.  Evidently, the same planar sector can yield a cone with apex that points up or down, depending on whether the surface of the sector appears on the outside or inside of the cone.

Note the two important limiting cases: 
\begin{enumerate}
\item If  $\alpha \gg 1$, then $\sin\theta\to 0$ and $\theta \to 0\  \hbox{or} \ \pi$.  Here the cone becomes narrow and approaches a cylinder with open end pointing down or up, respectively (see below  for more detailed discussion).
\item If $\alpha \to 1^+$, then $\sin\theta \to 1^-$ and $\theta \to \pi/2$. Here the cone becomes flat and the sector approaches the full plane.
\end{enumerate}
For any $\alpha >1$, the cone has a sharp  apex where the curvature is singular; this singularity disappears only for the special value $\theta = \pi/2$ or equivalently $\alpha = 1$.  On the smooth conical surface,  the curvature $\kappa_1$ vanishes along the radial direction and the curvature  along the azimuthal direction is $\kappa_2 =\mp \cot\theta /r= \mp\sqrt{\alpha^2-1}/r$.  Hence the Gaussian curvature $K = \kappa_1\kappa_2$ vanishes but the mean curvature  $H = \frac{1}{2} (\kappa _1 + \kappa_2) = -\frac{1}{2} \bm\nabla\cdot \hat{\bm n}$ is generally nonzero.  These results follow from the material in Secs.~IV and V of Ref.~\cite{Kami02}.

\subsection{Conformal map, complex potential and vortex dynamics}
In this way, the behavior of a quantized vortex on the surface of a cone becomes equivalent to that of a two-dimensional vortex on a planar sector with opening angle $2\pi/\alpha$ and periodic boundary conditions on the 
two straight sides. 
Similar to the case of a cylinder and its equivalent infinite planar strip with periodic boundary conditions (see Refs.~\cite{Ho15,Guen17}, and Appendix~\ref{sec:cylinder}), the conformal transformation 
\beq
z=Z^{1/\alpha} \longleftrightarrow Z = z^\alpha
\eeq 
maps the whole plane with $Z =|Z|e^{i\phi}$ to the unbounded sector with $z = |z|e^{i\bar\phi}$.
Hence we have $Z = |Z|e^{i\phi}= |z|^\alpha e^{i\alpha \bar\phi}$, where $-\pi\le \phi\le \pi$  and $-\pi/\alpha\le \bar\phi\le \pi/\alpha$ (for an application to electrostatics, see Sec.~25.2 in Ref.~\cite{Rile06}). 
 The action of this map is illustrated in Fig.~\ref{fig:map_to_cone}.
Notice that for complex numbers the power function is defined through the logarithm function,  $z^\alpha\equiv e^{\alpha \ln z}$, so that a choice of branch cut of the logarithm different from the principal one will result in a different angular range for the variable $\bar\phi$.

On the infinite plane, the complex potential for a positive singly quantized vortex at position $Z_0$ is simply given by Eq.~\eqref{Fplane}.  The conformal transformation immediately gives the equivalent result for a positive vortex on an unbounded sector with apex angle $\theta = \arcsin(1/\alpha)$: 
\begin{equation}\label{Fcone}
F_{\rm cone}(z) = \ln\left(z^\alpha - z_0^\alpha\right).
\end{equation}
Note that for $\alpha = n$ (an integer), this result is simply the sum of contributions for the original vortex and its $n-1$ images equally spaced on a circle of radius $|z_0|$. In terms of the variables $(r,\phi)$ on the cone, the corresponding stream function reads
\beq\label{chi_cone_r_phi}
\chi_{\rm cone}(\br,\br_0) =\textstyle{\frac{1}{2}}\ln\left[r^{2\alpha}-2 r^\alpha r_0^\alpha \cos(\phi-\phi_0)+r_0^{2\alpha}\right].
\eeq
For a general $F(z)$ that includes multiple vortices and perhaps boundaries, it is not hard to show that the motion of a positive vortex at $z_0 = r_0 e^{i\phi_0}$ is 
\begin{equation}\label{vel}
\dot{y}_0 + i\dot{x}_0 =\frac{\hbar}{M}\left[ \frac{dF}{dz} -\frac{1}{z-z_0}\right]_{z\to z_0},
\end{equation}
where the last term subtracts the (singular) circulating flow from the vortex itself.
 Given the complex potential in Eq.~\eqref{Fcone} for an unbounded sector (and hence for the surface of an infinite cone), an elementary calculation gives the result  
 \begin{equation}
\dot{y}_0+ i \dot{x}_0 = \frac{(\alpha-1) \hbar}{2M z_0}
\end{equation}
or, equivalently, in vector form,
\begin{equation}\label{linvel}
\dot{\bm r}_0 = \frac{(\alpha-1)\hbar}{2M r_0}\,\hat{\bm n}\times \hat{ \bm r}_0,
\end{equation}
where $x_0 = r_0\cos\bar\phi_0$ and $y_0 = r_0\sin\bar\phi_0$.  If $\alpha \to 1$, the vortex becomes stationary, as expected because the sector then covers the whole plane and the equivalent cone becomes flat.

Note that the motion is purely azimuthal, so that the time taken to complete one cycle on the sector is 
$$t_0 = \frac{2\pi r_0}{\alpha\,|\dot{\bm r}_0|} =  \frac{4\pi Mr_0^2}{\alpha(\alpha-1)\hbar}.$$
The frequency of the cyclic vortex motion around  the cone is $ 1/t_0$, and correspondingly the angular frequency around the cone is  
\begin{equation}\label{dotphi0}
\dot{\phi}_0 = \frac{2\pi}{t_0} = \frac{\alpha(\alpha-1)\hbar}{2Mr_0^2}, 
\end{equation}
expressed in terms of the radial distance $r_0$ along the surface from the apex of the cone. 
 For many purposes, the more relevant distance is $r_\perp =r_0\sin\theta = r_0/\alpha$, yielding 
\begin{equation}\label{angvel}
\dot{\phi}_0 = \left(1-\frac{1}{\alpha}\right)\frac{\hbar}{2M r_\perp^2}.
\end{equation}

In the limit $\alpha \gg 1$, the apex angle $\theta = \arcsin(1/\alpha )\approx 1/\alpha$ becomes small, and the cone locally  approaches a cylinder. In this limit Eq.~(\ref{angvel}) correctly reduces to the quantized result $\dot{\phi}_0 = \hbar/(2M r_\perp^2)$ that we found for an infinite cylinder of radius $r_\perp$ in  Ref.~\cite{Guen17}.

It is also instructive to examine the limiting behavior for $\alpha \gg 1$ of the conformal transformation $Z = z^\alpha$ from a planar sector to the whole plane.  Specifically, we now show that this transformation locally becomes that for the conformal transformation from a strip with periodic boundary conditions to the whole plane, which is discussed in detail in Appendix~\ref{sec:cylinder}.
Indeed, consider a point on the sector with  $|z|\approx 1$.  Let  $z =|z| e^{i\bar\phi}$ with $|z| = 1 + \epsilon$ and $-\pi/\alpha < \bar\phi < \pi/\alpha$.   
Here $z^\alpha = e^{\alpha \ln z} = e^{\alpha [\ln(1+\epsilon) + i\bar\phi]} \approx e^{\alpha(\epsilon + i \bar\phi)}$.  The rescaling $\bar z = \alpha\bar\phi - i\alpha \epsilon = \bar x + i\bar y$  then gives the desired result $z^\alpha \approx e^{i\bar{z}}$.  This mapping naturally associates the angle $\bar\phi$ on the sector with the angle $\phi=\alpha \bar\phi$ on the cylinder.  

Finally, let us take a close look at the local flow around a vortex core located at $\br_0$. The stream function Eq.~\eqref{chi_cone_r_phi} may be expanded by introducing $\delta r=r-r_0$ and $\delta\phi=\phi-\phi_0$, to obtain
\beq\label{chi_cone_r_phi_approx}
\chi_{\rm cone}(\br,\br_0)\approx\textstyle{\frac{1}{2}}\ln\left[\alpha^2r_0^{2\alpha-2}
\left(\delta r^2+r_\perp^2\delta\phi^2\right)\right],
\eeq
which shows explicitly that, very close to the vortex core, the stream function is constant for circles defined by the constant squared distance $ds^2 = \delta r^2+r_\perp^2\delta\phi^2$. This result is expected, since vortex cores on the plane are circular, and conformal transformations preserve the shapes of infinitesimal objects. Circular streamlines are indeed visible in the vicinity of the cores in both Figs.~\ref{fig:sector} and \ref{fig:truncated_cone}.

\section{Energy of two vortices}
\label{sec:energy}
 
In the present hydrodynamic model, the total energy $E_{\rm tot}$ of two vortices on a cone is purely kinetic:
\beq
E_{\rm tot} =\textstyle{ \frac{1}{2}} nM \int d^2 r\,v^2,
\eeq
 where $n$ is the two-dimensional number density and $\bm v = \hat{\bm n}\times \bm \nabla \chi_{\rm tot}$ is the total velocity.  Here, $\chi_{\rm tot} = q_1\chi_1 + q_2\chi_2$ with $\chi_j =\chi_{\rm cone}(\bm r,\bm r_j)$, and $q_j$ the charge of vortex $j$. 
The total energy may be written as $E_{\rm tot}=E_1+E_{12}+E_2$, where $E_{12}$ is the interaction energy between the two vortices, and $E_1$ and $E_2$ are the corresponding self-energies.
 
 \subsection{Interaction energy}
 
The two cross terms in $E_{\rm tot}$ yield the interaction energy of two vortices   
\beq
E_{12} =q_1q_2  \frac{n\hbar^2}{M} \int d^2 r\, \bm \nabla \chi_1\cdot \bm \nabla\chi_2.
\eeq
Use the two-dimensional divergence theorem to rewrite the integral as 
\begin{equation}\label{Eint}
I_{12} = \oint _{\cal C} dl \chi_1\,\hat{\bm \nu}\cdot \bm \nabla \chi_2 -\int d^2 r \chi_1\nabla^2\chi_2,
\end{equation}
where $\hat{\bm \nu}$ is the outward unit normal in the surface to the various boundaries.
The second term immediately gives $-2\pi \chi_{12}= -2\pi\chi_{\rm cone}(\bm r_1,\bm r_2)$ because $\nabla^2\chi_2 = 2\pi\delta^{(2)}(\bm r-\bm r_2)$.

The first term of Eq.~(\ref{Eint}) is a line integral around the boundary of the cone.  It is natural to take two circles: one  at radial distance $\epsilon \ll r_j$  since the apex of the cone is a singular region, and the other at  $R\gg r_j$  because the overall integral is log divergent.

For small $r=\epsilon$, the unit normal vector is $\hat{\bm \nu} = - \hat{\bm r}$.  The relevant derivative becomes $\partial\chi_2/\partial r |_\epsilon \approx -\alpha\epsilon^{\alpha -1}\cos(\phi-\phi_2)/r_2^\alpha$.  The circumference is $2\pi \epsilon/\alpha$, so that this contribution vanishes for $\epsilon\to 0$ (note that this line integral  also vanishes because of  the cos factor).

On the  large circle the stream function becomes $\chi_2 \approx \ln r^\alpha$ so that $\partial \chi_2/\partial r\approx \alpha/r$.  Here the unit vector  $\hat{\bm \nu}$ is simply $\hat{\bm r}$ and the circumference is $2\pi R/\alpha$, so that this term contributes $2\pi \ln R^\alpha$.  Consequently, we find 
\begin{equation}\label{int}
E_{12} =-q_1q_2 \frac{\pi n \hbar^2}{M} \ln\left(\frac{r_1^{2\alpha} - 2r_1^\alpha r_2^\alpha\cos\phi_{12} + r_2^{2\alpha}}{R^{2\alpha}}\right),
\end{equation}
where $\phi_{12} = \phi_1-\phi_2$.  Note that  the argument of the logarithm is dimensionless, as it must be.
This function is periodic in the relative angular displacement $\phi_1-\phi_2$, but the dependence on the radial position  ($0< r_j<\infty$) is more complicated, with the power laws involving the parameter $\alpha$.  This behavior reflects the loss of translation symmetry along the radial axis of the cone because the cone's tip serves as the origin of the spherical polar coordinates ($r,\phi$) with fixed $\theta$.

\subsection{Self-energy}
Despite the loss of translational symmetry, the self-energy for a single vortex at position $(r_1,\phi_1)$ on a large cone of radial dimension $R\gg r_1$  is scarcely more intricate than for a plane or a cylinder.  
As for the interaction energy, we again integrate by parts 
\begin{equation}\label{E1}
E_1 = \frac{n\hbar^2}{2M} \int d^2 r\, \bm\nabla \cdot \left(\chi_1\bm\nabla \chi_1\right) -  \frac{n\hbar^2}{2M} \int d^2 r\,
\chi_1 \nabla^2 \chi_1,
\end{equation}
where the integral is over the surface of the bounded cone with $r = R \gg r_1$  excluding a small circle of radius $\xi$ around the vortex core (and a small circle around the apex of the cone, which is irrelevant here).  This  exclusion near the vortex means that the second term in Eq.~(\ref{E1}) never contributes. In addition, the first term can be evaluated readily with the two-dimensional divergence theorem
 \begin{equation}
E_1 =  \frac{n\hbar^2}{2M}\oint dl\, \chi_1 \,\hat{\bm \nu}\cdot \bm\nabla\chi_1,
\end{equation}
where the integral is over all boundaries on the surface and $\hat{\bm \nu}$ is the outward unit normal in the surface to the boundaries of the original area (here the large circle at $r = R$ and the small circle around the vortex core).

Start with the large circle at $r = R$, where  $\hat{ \bm \nu}$ is just the unit vector $\hat{\bm r}$. For $r\gg r_1$, the stream function (\ref{chi_cone_r_phi}) simplifies to $\chi_{\rm cone} \approx \ln r^\alpha$, and a straightforward calculation gives the contribution $(\pi n \hbar^2/M) \ln R^\alpha$.

Next, consider the small circle around the vortex core, where Eq.~(\ref{chi_cone_r_phi_approx}) shows that the flow is axisymmetric around the vortex.  Define $u^2 = \delta r^2 + r_\perp^2 \delta\phi^2$ so that $u$  a local radial variable centered on the vortex.  Hence the local stream function (\ref{chi_cone_r_phi_approx})  becomes  $\chi_{\rm cone }  \approx \ln \left( \alpha\,r_1^{\alpha-1}\,u\right)$ and the local outward normal is $\hat{\bm \nu } = -\hat{\bm u}$.  It is not hard to find the corresponding contribution to the self-energy  $-\left(\pi n\hbar^2/M\right)\ln\left(\alpha\,r_1^{\alpha-1} \xi\right)$.

Together, these two contributions yield the relevant self-energy of a singly quantized vortex at $r_1,\phi_1$ on a large truncated cone with bounding radial coordinate $r = R\gg r_1$ (correspondingly $R_\perp = R\sin\theta = R/\alpha$)
\begin{equation}\label{E11}
E_{1}\approx \frac{\pi n \hbar^2}{M}\ln\left(\frac{R^\alpha}{\alpha \xi r_1^{\alpha - 1}} \right)\approx  \frac{\pi n \hbar^2}{M}\alpha \ln\left(\frac{R}{\xi}\right),
\end{equation}
where the second expression keeps only the leading logarithmic term for $R/\xi \gg 1$.
If $\alpha = 1$, the first  expression reduces to the familiar self-energy for a vortex on a plane.  Otherwise,  $E_1$ explicitly involves the radial position  $r_1$ of the vortex on the cone, but only as an additive logarithmic constant.

\subsection{Total energy of vortex dipole on a cone}

A combination of the various terms yields the total energy  $E_{\rm tot} = E_1 + E_2 + E_{12}$ for the vortex dipole
\begin{equation}\label{Etot}
E_{\rm tot} = \frac{\pi n \hbar^2}{M} \ln\left(\frac{r_1^{2\alpha} - 2r_1^\alpha r_2^\alpha\cos\phi_{12} + r_2^{2\alpha}}{\alpha^2 \xi^2 r_1^{\alpha -1}r_2^{\alpha -1}}\right).
\end{equation}
As expected, this is independent of $R$ since the total charge vanishes.  Any dynamical motion of a vortex dipole must maintain constant total energy.

\begin{figure}
\begin{center}
\includegraphics[width=\linewidth]{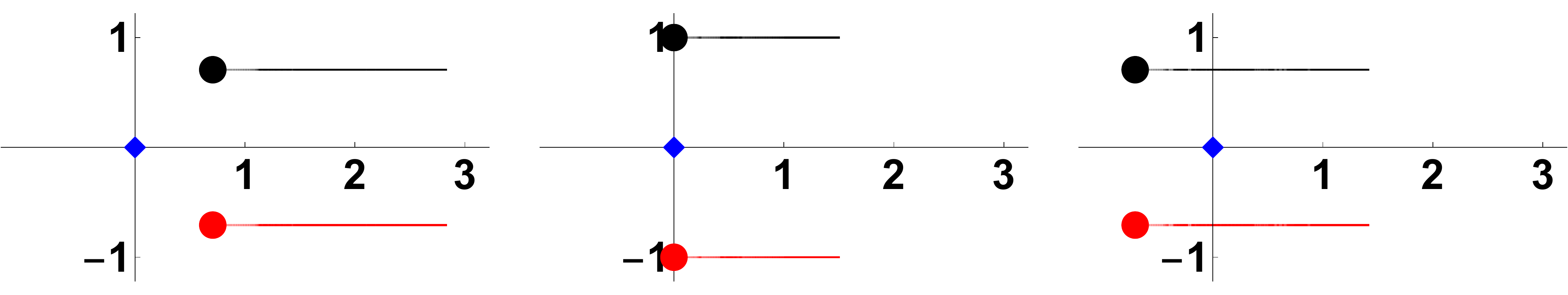}
\caption{\label{fig:+-_vortex_dynamics}
Trajectories of a {\bf vortex dipole} on an unbounded cone with $\alpha=3$. This image depicts directly $r_j e^{i\phi_j}$, where $r_j$ and $\phi_j$ ($j=1,2$) are the radial and angular positions of the two vortices on the cone, as given by Eqs.~\eqref{eq:dynamics_factorized_rDot} and \eqref{eq:dynamics_factorized_phiDot}. In each of the panels, the blue diamond denotes the tip of the cone. The positive vortex 1 and the negative vortex 2 start at $t=0$ from the black and red dots, respectively, which are located at unit radius and angles $\pm\pi/4$ (left), $\pm\pi/2$ (center), and $\pm3\pi/4$ (right), respectively, and evolve freely on the surface until $t_{\rm max}=M r_0^2/\hbar$. 
As time passes, the color of the trajectory gets increasingly dark.  In the right figure, the vortices initially move toward the apex, pass over the shoulder at $\phi_0 =\pm\pi/2$, and then move away from the apex.
}
\end{center}
\end{figure}

This expression becomes particularly simple for the symmetric initial condition $r_1=r_2 = r_0$ and $\phi_1 =-\phi_2 = \phi_0$, which yields 
\begin{eqnarray}\label{Edip}
E_{\rm tot} &=& \frac{\pi n\hbar^2 }{M} \ln\left[\frac{2r_0 ^2\left(1-\cos(2\phi_0)\right)}{\alpha^2 \xi^2}\right] \nonumber \\  &=& \frac{2\pi n\hbar^2 }{M} \ln\left(\frac{2r_0 \sin\phi_0}{\alpha \xi}\right).
\end{eqnarray}
To be very specific, any allowed motion of this special vortex dipole must conserve the product $r_0\sin\phi_0$.
If $q_1 =1$ and $q_2= -1$ and $0<\phi_0<\pi/2$, the flow through the center of the vortex dipole is away from the apex and the dipole will move in the same direction, with $r_0$ increasing. 
 In this case, $\sin\phi_0$ must correspondingly decrease and this process can continue indefinitely.  In contrast, if $\pi/2<\phi_0<\pi$,  then $r_0$ starts to decrease and the dipole moves toward the apex.  This  motion must eventually reach a turning point because $\sin\phi_0 $ cannot exceed 1.  It then turns around and moves away from the apex, as seen in the last part of Fig.~\ref{fig:+-_vortex_dynamics}.  Note that none of this analysis  gives any information about the local speed along the trajectory.

To interpret this expression in another way, note that for this special initial condition, the three-dimensional vector separation between the two vortices is $\bm r_1-\bm r_2 = 2r_0 \sin\theta\,\sin\phi_0 \,\hat{\bm y} = (2r_0 \sin\phi_0/\alpha)\, \hat{\bm y}$ and lies along $\hat{\bm y}$.  Hence the constancy of $E_{\rm tot}$ implies that $(\bm r_1-\bm r_2)$ must remain fixed and parallel to $\hat{\bm y}$.

\section{Dynamics of two vortices on a cone}\label{dyn}

On an unbounded plane and on an infinite cylinder~\cite{Guen17}, a vortex dipole moves uniformly  perpendicular to the line between their centers and in the same direction as the flow between them.  In both cases, the net  vortex-charge neutrality means that the relative vector $\bm r_{12} = \bm r_1-\bm r_2$ is a constant of the motion.  Ultimately, this behavior reflects the translational and rotational invariance of these surfaces.

Unlike the case of motion on a cylinder, a single vortex at $\bm r_1$ on a cone has a self-induced motion $\bm{\dot}{\bm r}_1 $  that depends on its specific position.  When there is an additional vortex at $\bm r_2$, the translational velocity of the first vortex  gains an added contribution arising  from the hydrodynamic flow from vortex 2 evaluated at $\bm r_1$, namely  $\bm v_2(\bm r_1) =q_2  (\hbar/M) \hat{\bm n} \times \bm \nabla_1 \chi_{\rm cone}(\bm r_1,\bm r_2)$.   

The details are straightforward.  For example, the combined translational velocity of vortex 1 has the explicit form
\begin{eqnarray}\label{eq:dynamics}
\bm{\dot}{\bm r}_1 &=&  \frac{\hbar q_1}{2Mr_1} (\alpha -1) \hat{\bm \phi}_1 
  \\
&+& \frac{\hbar q_2\,\alpha \,r_1^\alpha}{Mr_1}\frac{\left[r_1^\alpha - r_2^\alpha \cos \phi_{12}\right]\hat{\bm \phi}_1 -r_2^\alpha \sin\phi_{12}\,\hat{\bm r}_1 }{\rho^2},\nonumber
\end{eqnarray}
where $\rho^2 = \left(r_1^{2\alpha} -2r_1^\alpha r_2^\alpha \cos\phi_{12} + r_2^{2\alpha}\right)$, and $\phi_{12}=\phi_1-\phi_2$.
A similar expression holds for $\bm{\dot}{\bm r}_2$.  Note that the unit vectors ($\hat{\bm r}_j,\hat{\bm \phi}_j$)  here vary locally and differ for the two vortices.  The first term on the right side arises from the self-induced motion and is purely azimuthal, depending on the local position and the sign of the charge $q_1$.  In contrast,  the second term is the motion induced by the hydrodynamic flow of second vortex.  It depends on the positions of both vortices and on the charge $q_2$.

\begin{figure}
\begin{center}
\includegraphics[width=\linewidth]{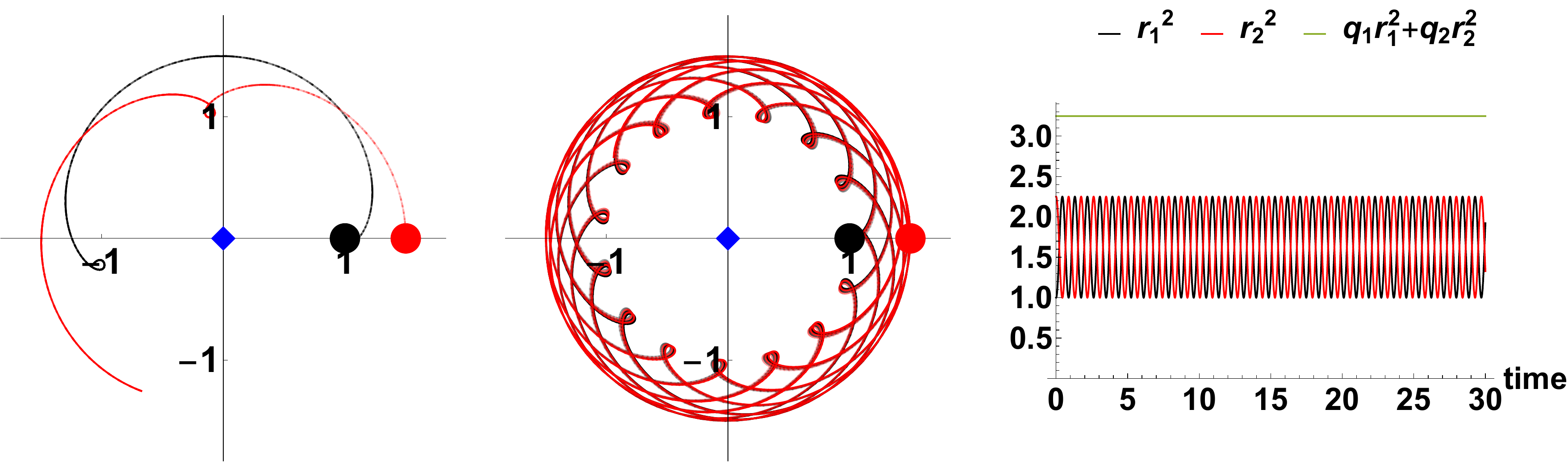}\\\vspace{0.4cm}
\includegraphics[width=\linewidth]{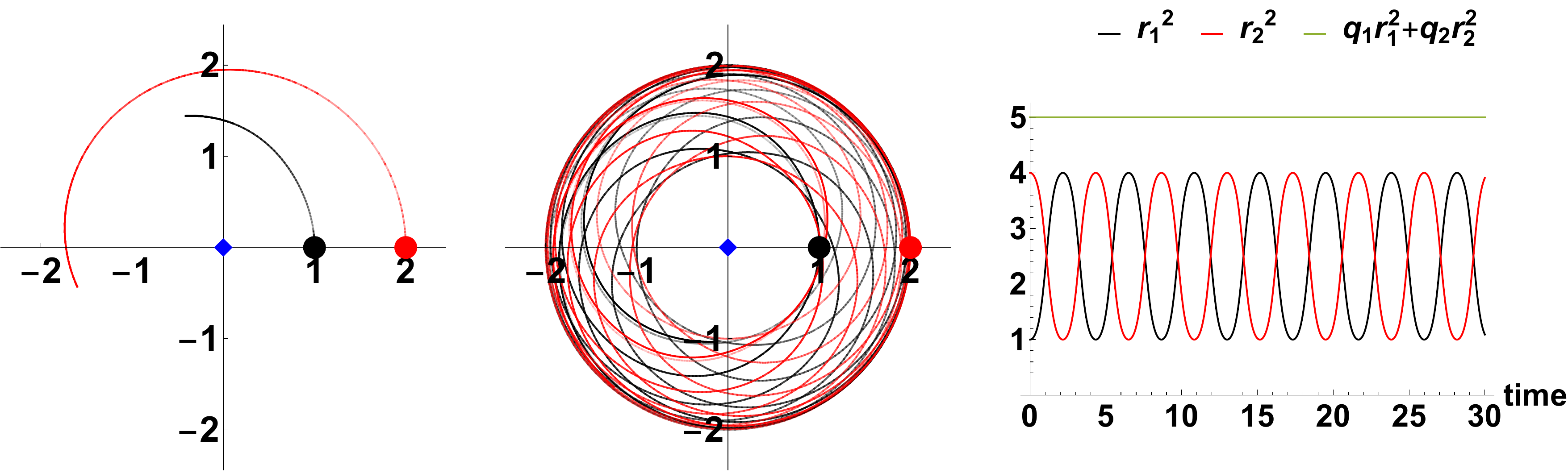}\\\vspace{0.4cm}
\includegraphics[width=\linewidth]{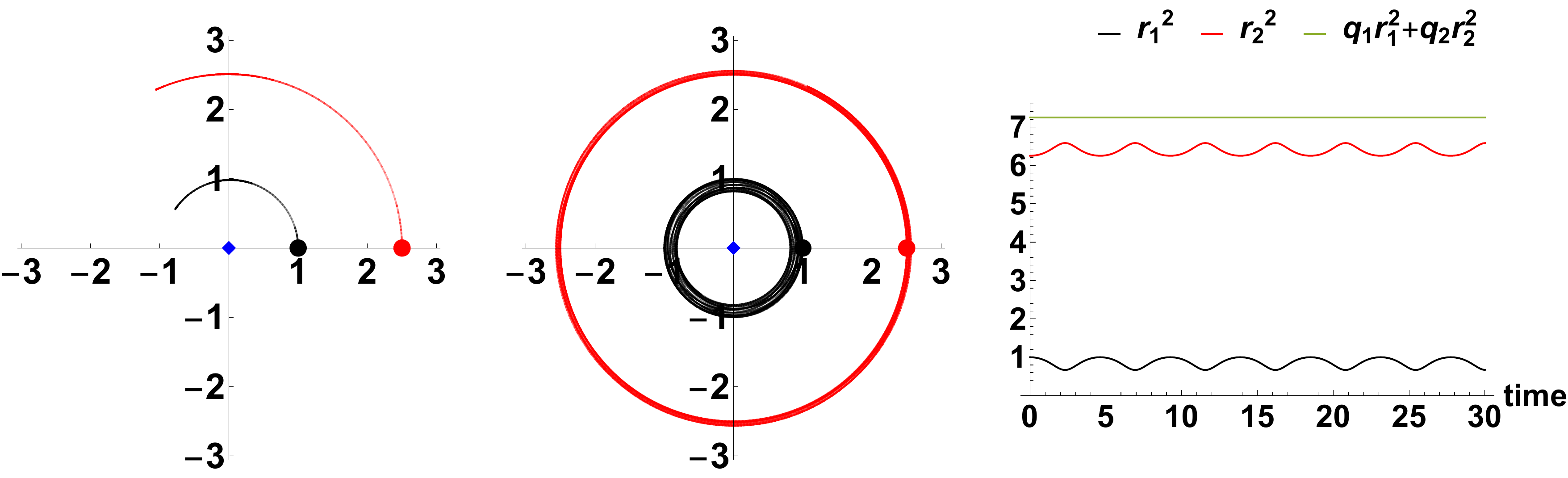}
\caption{\label{fig:++_vortex_dynamics}
Trajectories of a {\bf pair of positive vortices}  on an unbounded cone with $\alpha=3$.  Initial conditions are $r_1 = 1$ and $r_2 = 1.5 $ (top row), 2.0 (middle row) and 2.5 (bottom row), with $\phi_1=\phi_2 = 0$ initially.
Left: the evolution stops at $t_{\rm max}=M r_0^2/\hbar$. Center: the evolution continues until $t_{\rm max}=30M r_0^2/\hbar$.  Right figures show $r_1^2$ and $r_2^2$, along with their sum, which indeed remains constant during the whole evolution.
}
\end{center}
\end{figure}

To integrate these equations, recall that ${\bm r}=r\hat{\bm r}$, so that the time derivative of $\bm r$ yields 
\begin{equation}
\dot{\bm r} = \dot r\, \hat{\bm r} + r\sin\theta\, \dot\phi\, \hat{\bm \phi}.
\end{equation}
 Equation~\eqref{eq:dynamics} gives the two coupled equations
\begin{eqnarray}\label{eq:dynamics_factorized_rDot}
\dot r_1 &=&\frac{-\hbar q_2 \alpha r_1^\alpha r_2^\alpha\sin\phi_{12}}{Mr_1 \rho^2}\\ 
\label{eq:dynamics_factorized_phiDot}
 r_1\sin\theta\dot\phi_1 &=& \frac{\hbar q_1(\alpha-1)}{2Mr_1} + \frac{\hbar q_2 \alpha r_1^\alpha }{Mr_1}\frac{r_1^\alpha -r_2^\alpha\cos\phi_{12}}{\rho^2}.
\end{eqnarray}
Similar equations hold for the second vortex.

These coupled equations have a remarkably simple first integral:  The antisymmetry of $\sin\phi_{12}$ for $1\leftrightarrow 2$ shows that $q_1r_1\dot
r_1 + q_2 r_2 \dot r_2 =0$.  Hence the quantity $q_1 r_1 ^2 + q_2 r_2^2 $ is conserved.  In combination with the conservation of $E_{\rm tot}$, the problem effectively reduces to two variables and becomes relatively straightforward.

For two vortices with opposite signs (a vortex dipole with $q_1q_2 = -1$) the quantity $r_1^2- r_2^2$ remains fixed, allowing each correlated variable to become large.  Figure \ref{fig:+-_vortex_dynamics} illustrates such behavior for symmetric initial conditions.  In contrast, for two vortices with the same sign ($q_1q_2 = 1$), the quantity $r_1^2 + r_2^2$ is fixed, so that the motion of the two vortices remains localized, as seen in Fig.~\ref{fig:++_vortex_dynamics} for several symmetric  initial conditions.

A few cases  are simple to describe. The first is a vortex dipole with $r_1=r_2=r_0$ and $\phi_1=-\phi_2=\phi_0$. The four coupled equations now yield the single pair 
\begin{equation}
\dot r_0 = \frac{\hbar q_1 \alpha \cot\phi_0}{2Mr_0}\quad\hbox{and}\quad \dot\phi_0 = -\frac{\hbar q_1 \alpha }{2Mr_0^2}.
\end{equation}
Since $\dot r_0 + r_0\cot\phi_0\, \dot\phi_0 =0$, these equations are equivalent to the condition  that $r_0\sin\phi_0 $ is constant. The corresponding dynamics is shown in Fig.~\ref{fig:+-_vortex_dynamics}.  Note that this condition ensures the conservation of total energy as seen in Eq.~(\ref{Edip}).

The other example is a pair of equal-charge vortices ($q_1=q_2=q_0$) with initial conditions along the same radial direction $r_1\neq r_2$ and $\phi_2=\phi_1=0$. 
Figure \ref{fig:++_vortex_dynamics} shows the resulting dynamics for three cases:  $r_2/r_1 = 1.5, 2.0$, and 2.5. 
In the first, the interaction of the two vortices dominates the dynamics, whereas for the last, the vortices move essentially independently under the influence of the cone.  In each case, we show explicitly that $r_1^2 + r_2^2$ remains fixed.

\section{Cones with finite boundaries}
\label{sec:finiteCones}

This section uses the method of images on the bounded sector to determine the appropriate stream function on the bounded cone and the dynamical motion of a vortex with this geometry.

\subsection{Finite cone with  outer radius  $R_2$}

\begin{figure}[t]
  \begin{center}
  \includegraphics[width=2in]{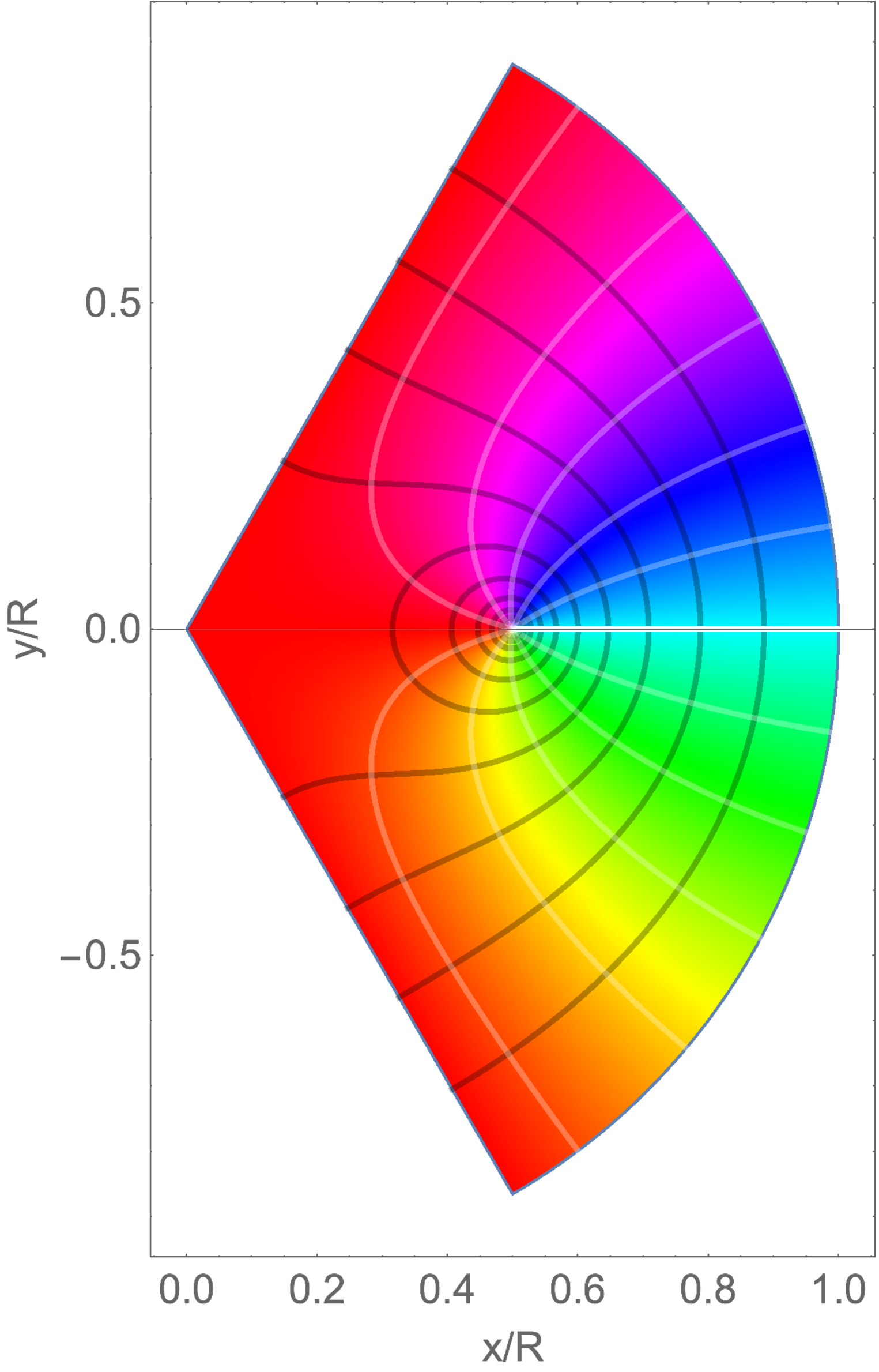}
  \caption{\label{fig:sector}
  Streamlines (black) and lines of constant phase (white) for positive unit vortex at $z_0 = 0.5\, R_2$ on a sector with $\alpha = 3$,  fixed outer radius $R_2$, and periodic boundary conditions (equivalent to a truncated cone of radius $R_2$).  These lines arise from the real and imaginary parts of the complex function $F_{\rm bounded}(z)$ in Eq.~(\ref{trcone}).
The Hue coloring of the plot denotes the phase of the wave function.
}
  \end{center}
 \end{figure}

\begin{figure}[t]
  \begin{center}
  \includegraphics[width=3in]{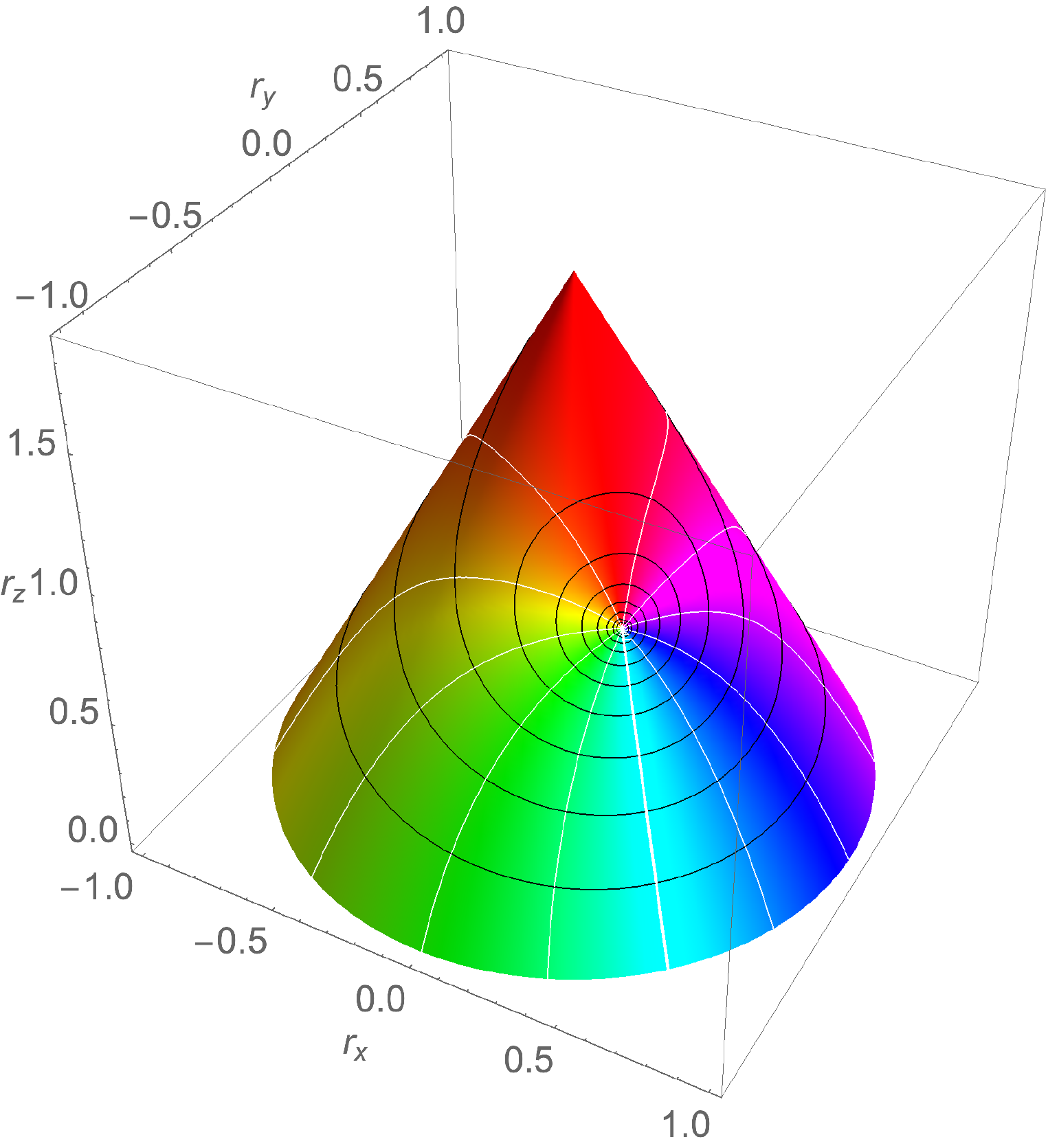}
  \caption{\label{fig:truncated_cone}
  Streamlines (black) and lines of constant phase (white) for positive unit vortex  on a truncated cone with $\alpha = 2.1$, opening angle $\theta \approx 0.5$ and $r_0 = 0.5\, R_2$.  These lines arise from the real and imaginary parts of the complex function $F_{\rm bounded}(z)$ in Eq.~(\ref{trcone}).}
  \end{center}
 \end{figure}

For a finite sector  of radius $R_2$ with $0 < r < R_2$, the method of images gives the intuitive result (compare Ref.~\cite{Guen17})
\begin{equation}\label{trcone}
F_{\rm bounded}(z) = \ln\left(\frac{z^\alpha - z_0^\alpha}{z^\alpha -(R_2^2/z_0^*)^\alpha}\right).
\end{equation}
With this complex potential, it is straightforward to visualize the streamlines (black) and lines of constant phase (white) for a positive vortex, as shown in Fig.~\ref{fig:sector}.
It is also not difficult to project a similar phase pattern onto the surface of an equivalent three-dimensional truncated cone, as shown in Fig.~\ref{fig:truncated_cone}.

 The precessional velocity of a positive vortex at $z_0 = r_0 e^{i\phi_0}$ on the bounded cone directly follows from Eq.~(\ref{vel}):
 \begin{equation}
\dot{\bm r}_0 = \frac{\hbar}{Mr_0}\left(\frac{\alpha-1}{2} + \frac{\alpha r_0^{2\alpha}}{R_2^{2\alpha} -r_0^{2\alpha}}\right) \hat{\bm \phi}_0,
\end{equation}
where the first term is that for an unbounded cone and the second reflects the presence of the outer boundary.
For $\alpha \to 1$ (the planar case), this result reproduces the standard expression from classical hydrodynamics [see, for example, Eq.~(B7) in Ref.~\cite{Guen17}].
For $r_0\ll R_2$, the second term vanishes, reproducing the result for an unbounded cone.  In contrast,  the second term predominates for $r_0 \approx R_2^-$, when the precessional velocity increases rapidly.

 \subsection{Truncated unbounded cone with inner radius $R_1$}
 
 A similar formalism describes the motion of a positive vortex on an unbounded truncated cone with $R_1 < r < \infty$ where $r$ again is the radial coordinate measured along the surface of the cone.  The complex potential is now
 \begin{equation}
F_{\rm trunc}(z ) = \ln \left(\frac{z^\alpha - z_0^\alpha}{z^\alpha -(R_1^2/z_0^*)^\alpha}\right) + \ln z^\alpha.
\end{equation}
Here the last term reflects an additional positive quantized vortex at the origin ensuring that the net circulation around the inner boundary vanishes.

Equation (\ref{vel}) readily yields the corresponding precessional velocity 
\begin{equation}
\dot{\bm r}_0 = \frac{\hbar}{Mr_0}\left(\frac{\alpha-1}{2} - \frac{\alpha R_1^{2\alpha}}{ r_0^{2\alpha} - R_1^{2\alpha}}\right) \hat{\bm \phi}_0.
\end{equation}
For $\alpha \to 1$ (the planar case), this result reproduces the classical expression.
Here, the second term (reflecting the combined effect of the image and the vortex at the origin) now induces a negative motion when $r_0\approx R_1^+$.

\subsection{Truncated bounded cone with $R_1< r < R_2$} 
As seen in our earlier works on the annulus (Ref.~\cite{Fett67}) and on the cylinder (Ref.~\cite{Guen17}), the inclusion of two boundaries leads to a doubly infinite set of images.  
 Equation (B3) in Ref.~\cite{Guen17} gives the complex potential for a vortex  in a planar  annulus with $R_1< r_0 < R_2$.  The corresponding complex potential for  a vortex at $z_0 = r_0e^{i\phi_0}$ in a bounded planar sector with opening angle $2\pi/\alpha$ follows immediately with the conformal transformation $Z = z^\alpha$:
 \begin{equation}
F_{\rm bounded}(z) = \ln\left[\frac{\vartheta_1[-(i/2)\ln(z^\alpha/z_0^\alpha), (R_1/R_2)^\alpha]}{\vartheta_1[-(i/2)\ln(z^\alpha/z_0'^\alpha), (R_1/R_2)^\alpha]}\right].
\end{equation}
Here $z_0' = R_2^2/z_0^* = (R_2^2/r_0)e^{i\phi_0}$ is the complex  image  of $z_0$ with respect to the outer boundary.  Note that $z_0'$ lies  on the surface of the cone extended beyond the larger boundary.

Equation (\ref{vel}) again yields the translational velocity of the vortex $\dot{\bm r}_0 = v_0\, \hat{\bm \phi}_0$, where 
\begin{equation}\label{dotr}
v_0 = \frac{\hbar}{2Mr_0}\left[ i\alpha\,\frac{\vartheta_1'[-i\alpha \ln(r_0/R_2),(R_1/R_2)^\alpha]}{\vartheta_1[-i\alpha \ln(r_0/R_2),(R_1/R_2)^\alpha]}
-1\right]
\end{equation}
 defines the velocity $v_0$ (it is real, despite the explicit appearance of $i$).
The vortex moves uniformly around the bounded cone at fixed $r_0$.  In the limit $\alpha\to 1$, this expression properly reduces to that for a planar annulus, given in Eq.~(B4) of Ref.~\cite{Guen17}.   

\begin{figure}[t]
  \begin{center}
  \includegraphics[width=3.5in]{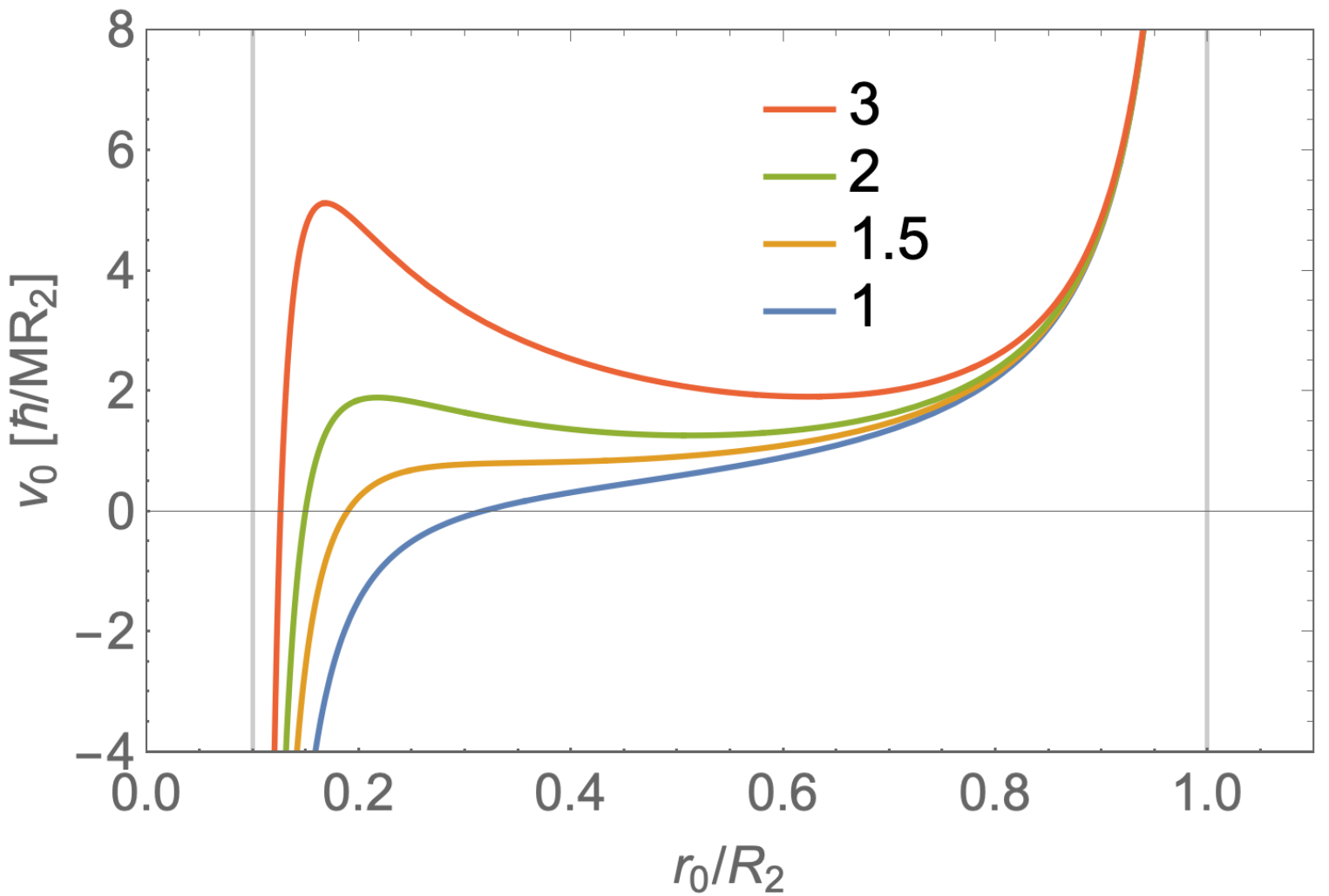}
  \caption{\label{fig:core_velocity}
  Translational velocity in the azimuthal direction $v_0  $ (in units of $\hbar/MR_2$) from Eq.~(\ref{dotr}) as function of vortex position $r_0$, for various values of $\alpha$ on a truncated bounded cone with $R_1/R_2 = 0.1$. The curve for $\alpha = 1$ is  similar to that in Fig.~7 of our earlier paper, Ref.~\cite{Guen17}.}
  \end{center}
 \end{figure}

Figure \ref{fig:core_velocity} shows $v_0(r_0)$ for typical values of the cone parameter $\alpha$. 
When $r_0$ approaches $R_2$, the outer image dominates and the vortex moves  rapidly in the positive direction.  In contrast, the behavior for $r_0 \to R_1$ depends on $\alpha$.  For not-too-large  $\alpha \gtrsim 1$,  the inner image dominates and the vortex simply  slows and then reverses, moving in the negative direction as $r_0\to R_1^+$.  For larger $\alpha$, however, the translational velocity initially increases with decreasing $r_0$  because of  the factor $r_0^{-1}$ in Eq.~(\ref{dotr}).  Eventually, the inner image dominates and the vortex then moves in the opposite direction.

For the special choice $r_0 =\sqrt{R_1R_2}$  (the geometric mean of the inner and outer radii), the identity $-i\vartheta_1'(-i\ln \sqrt q,q) =\vartheta_1(-i\ln\sqrt q,q)$ readily yields the simple result
\begin{equation}\label{v0}
v_0 = \frac{\hbar\,(\alpha-1)}{2M\sqrt{R_1R_2}},
\end{equation}
which is the same as Eq.~(\ref{linvel}) for an unbounded cone.  

\section{Outlook and conclusions}\label{sec:conclusions}
Our study of superfluid vortex dynamics on a cone with opening angle  $ \theta = \arcsin(1/\alpha)$ starts from a planar sector with opening angle $2\pi/\alpha \le 2\pi$, so that $\alpha \ge 1$.  For a complex variable $z$ on the sector, a  simple conformal transformation $Z(z)=z^\alpha$ maps this sector onto the full plane. 

 \subsection{Conjecture for  sector with $ 0 < \alpha < 1$}
 From this perspective of  such a planar sector, it is reasonable to ask what happens when $\alpha$ becomes smaller than $1$,   remaining real.  For definiteness, consider the rational number $\alpha = p/q$ where  $p$ and $q$ are co-prime numbers and $p<q$.  The sector now has opening angle $2\pi/\alpha = 2\pi q/p$ that exceeds $2\pi$ and thus extends beyond the single plane with an overlapping surface.  
The same conformal mapping $Z(z) = z^\alpha= z^{p/q}$ now has a Riemann surface with $q$ sheets forming a closed cyclic structure.  The simplest case is $\alpha = 1/2$, which leads to a two-fold closed structure for the Riemann surface. An irrational value of $\alpha$ leads instead to a Riemann surface which never folds onto itself.
 
 For such an overlapping sector, it appears that  the preceding formalism determining Eqs.~(\ref{linvel}) and (\ref{angvel}) for the induced vortex motion on the sector remains valid even for $\alpha <1$.  In particular, a positive  vortex should now move in the clockwise  direction because $\alpha-1$ is negative.
 
 The more difficult question is what, if any, three-dimensional surface corresponds to a sector of the plane with opening angle $2\pi/\alpha$ and $0<\alpha <1$.  Evidently this sector overlaps at least part of the plane more than once.  The inverse definition $\sin\theta = 1/\alpha$ requires that $\theta$ become complex with $\theta = \pi/2 + i \lambda$, and real $\sin\theta =\cosh\lambda=1/\alpha >1$.  Correspondingly, $\cos\theta = -i\sinh\lambda = -i\,{\rm sgn}\lambda\,\sqrt{1-\alpha^2}/\alpha$ is pure imaginary.  For this mapping, the definition $z = r\cos\theta$ yields an imaginary value for the three-dimensional vertical coordinate of what was a physical cone for $\alpha > 1$.  Hence there may well  be no simple three-dimensional  real physical surface corresponding to the planar sector with $\alpha < 1$.

\subsection{Comparison with previous work}
 
Vortex dynamics on a plane is an old subject dating back over a century~\cite{Lamb45}.  The plane is flat, so that the question of surface curvature never arises, and such studies are widely applied to quantized vortices in superfluid $^4$He~\cite{Donn91}.

Most of the earlier studies of vortices  on curved surfaces  focused on the total energy and its dependence on the location of the singularities of the relevant fields~\cite{Vite04,Turn10,Kami02,Ho15,Lube92,Mach89}.  This concentration on the energy yields useful thermodynamic information, but it ignores the very interesting question of the dynamical motion of such superfluid vortices.  

The generalization to vortices on curved surfaces has involved two distinct topological situations. Here and in Ref.~\cite{Guen17}, we considered   regions with multiply connected topology and associated winding numbers, specifically a cylinder and a cone.  In each case, we studied the associated dynamics of both a single vortex and two vortices (particularly a neutral  vortex dipole). 
The situation on a compact surface such as a sphere differs in that the total vortex charge must vanish~\cite{Vite04,Turn10,Lamb45}.

To our knowledge, the only other relevant discussion of vortex dynamics on a curved surface  is  Secs.~80 and 160 of Ref.~\cite{Lamb45}, which considers the case of vortices on a sphere.   In addition to the restriction to a vortex dipole with zero net charge, Lamb quotes the result for the translational velocity of a vortex dipole on a sphere, based on Kirchhoff's much earlier electrical studies of a thin spherical conducting layer (see Sec.~80 of Ref.~\cite{Lamb45}).

\subsection{Conclusions}

We have presented here a description of the dynamics of superfluid  vortices on the surface of a cone. We have demonstrated that single vortices precess uniformly around the symmetry axis of the cone (hence conserving self-energy), while configurations with two vortices remain quasiperiodic because of two conserved quantities.
As $\alpha$ varies between 1 and $+\infty$, the surface of a semi-infinite cone interpolates smoothly between the whole complex plane and an infinite thin cylinder. In the two limits, we have shown explicitly that the results presented here reproduce those well-known behaviors.
As a final remark, we note that our previous results on cylindrical surfaces may prove very useful to understand the physics at play in the very recent experiment that studied superfluid $^4$He adsorbed on carbon nanotubes \cite{Mena19,Nour19}.

\vspace{5mm} 
\section*{Acknowledgements}
The authors thank I.~Carusotto, N.-E.~Guenther and R.~B.~Laughlin  for insightful discussions. 
P.M. acknowledges support from Spanish MINECO (FIS2017-84114-C2-1-P) and the ``Ram\'on y Cajal'' program.   A.L.F. is grateful for the opportunity to visit S.~Stringari and the University of Trento in October 2017,  where part of this work was done.

\begin{figure*}
\begin{center}
\includegraphics[width=\linewidth]{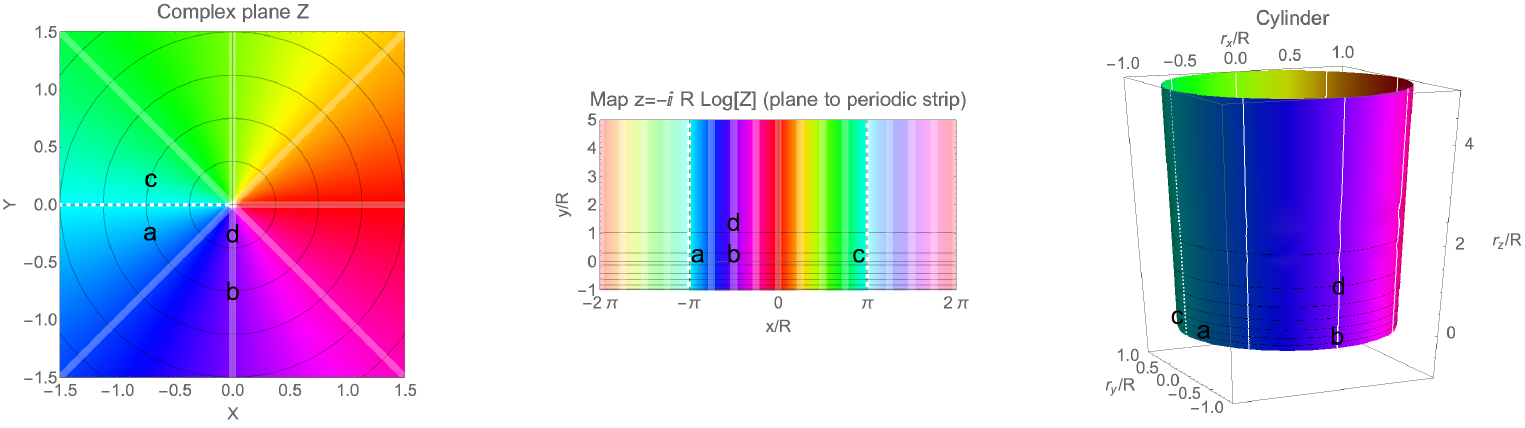}
\caption{\label{fig:map_to_cylinder}
{\bf Conformal map to a strip and cylinder.} The map $z=-i R \log{Z}$ sends the complex plane $Z$ (left) onto a strip of width $2\pi R$ (center). Different choices for the branch cut of the logarithm yield equivalent strips of equal width, displaced along $x$ by an arbitrary amount. Any of these strips may be folded onto a three-dimensional cylinder (right).  The map is conformal, so that lines of constant modulus $|Z|$ (black) and lines of constant phase $\phi$ (white) of the source variable on the plane $Z = |Z|e^{i\phi}$ are orthogonal in all panels. The Hue coloring of the maps corresponds to the phase of the coordinate on the source plane, $\phi={\rm Arg}(Z)$.}
  \end{center}
 \end{figure*}

\appendix
\section{Vortex motion on a cylinder}\label{sec:cylinder}
We here review the complex potential for a single vortex on a cylinder. The most straightforward approach \cite{Ho15} is to use a conformal map, which links the whole plane (with a coordinate $Z$) to a periodically-repeating strip of width $2\pi R$ (with a coordinate $z$):  
\begin{equation}\label{map_cylinder}
z=-i R \log{Z} \longleftrightarrow Z=e^{i z/R}.
\end{equation}\\
The action of this map is shown in detail in Fig.~\ref{fig:map_to_cylinder}. In particular, the negative real axis of the complex plane maps to the two infinite sides of the strip, located at coordinates $x = \pm \pi R$.  
The strip (shown in the central panel) features periodic boundary conditions, and therefore may be folded onto a three-dimensional cylinder (shown in the right panel). This procedure does not introduce any distortion, so that the transformation is conformal. The coordinates $\{r_x,r_y,r_z\}$ on the cylinder are obtained from the coordinates $\{x,y\}$ on the strip by the obvious relations $r_x=R \cos(x/R)$, $r_y=R \sin(x/R)$, and $r_z=y$.

The conformal map Eq.~\eqref{map_cylinder} may be combined with the complex potential on the plane, Eq.~\eqref{Fplane}, to obtain the corresponding potential on the strip and on the cylinder for a vortex at $z_0$:
\beq
F_{\rm str}(z)=F_{\rm cyl}(z)=\ln\left(e^{i z/R}-e^{i z_0/R}\right).
\eeq
Results for complex potentials, core velocities and interaction energies for finite and infinite cylinders were given in our previous work, Ref.~\cite{Guen17}. Therein was also a detailed discussion of the close connection of these results with those found earlier in Ref.~\cite{Fett67} for the topologically equivalent case of vortices on a planar annulus.

\end{document}